\documentclass{aipproc}
\layoutstyle{6x9}

\usepackage[english]{babel}
\usepackage{amsmath,amssymb,amsfonts,amstext}
\usepackage{hyperref}

\begin{document}
\title{Regge-model predictions for $K^+\Sigma$ photoproduction from the nucleon}

\author{P.~Vancraeyveld\footnote{Email address: Pieter.Vancraeyveld@UGent.be}}{
	address={Department of Physics and Astronomy, Ghent University, Proeftuinstraat 86, B-9000 Gent, Belgium}}
\author{L.~De~Cruz}{
	address={Department of Physics and Astronomy, Ghent University, Proeftuinstraat 86, B-9000 Gent, Belgium}}
\author{J.~Ryckebusch}{
	address={Department of Physics and Astronomy, Ghent University, Proeftuinstraat 86, B-9000 Gent, Belgium}}
\author{T.~Van~Cauteren}{
	address={Department of Physics and Astronomy, Ghent University, Proeftuinstraat 86, B-9000 Gent, Belgium}}

\begin{abstract}
We present Regge-model predictions for the $p(\gamma,K^+)\Sigma^0$ and $n(\gamma,K^+)\Sigma^-$ differential cross sections and photon-beam asymmetries in the resonance region. The reaction amplitude encompasses the exchange of $K^+(494)$ and $K^{\ast+}(892)$ Regge-trajectories, introducing a mere three free parameters. These are fitted to the available $p(\gamma,K^+)\Sigma^0$ data beyond the resonance region. The $n(\gamma,K^+)\Sigma^-$ amplitude is obtained from the $p(\gamma,K^+)\Sigma^0$ one through $SU(2)$ isospin symmetry considerations.
\end{abstract}

\classification{11.55.Jy, 12.40.Nn, 13.60.Le, 14.20.Gk}
\keywords      {$n(\gamma,K^+)\Sigma^-$ observables, Regge phenomenology}

\maketitle

Electromagnetic (EM) kaon production plays a key role in the ongoing theoretical and experimental efforts to resolve the missing resonance problem.  Despite the publication of a large body of high-quality $p(\gamma^{(\ast)},K)Y$ data in recent years, phenomenological analyses have not led to an unequivocal outcome. Disentangling the relevant resonant contributions is convoluted, because of the large number of competing resonances above the kaon production threshold. Moreover, the smooth energy dependence of the measured observables hints at a dominant role for background, i.e.\ non-resonant, processes. 

At sufficiently high energies and momentum transfers, the isobaric description breaks down as partons become the relevant degrees-of-freedom. In this energy region, the kaon production amplitude can be elegantly described within the Regge framework, characterized by the exchange of whole families of particles, instead of individual hadrons. Building upon the work of Guidal et al.~\cite{GuidalPhotoProdKandPi,PHDguidal}, we model the $p(\gamma,K^+)\Sigma^0$ amplitude by means of $K^+(494)$ and $K^{\ast+}(892)$ Regge-trajectory exchange in the $t$-channel~\cite{RPRsigma}. A gauge-invariant amplitude is obtained by adding the electric part of the nucleon $s$-channel Born diagram. The strong forward-peaked character of the differential cross section provides powerful support for this approach. In our implementation of the Regge model, the operatorial structure of the amplitudes is dictated by an effective Lagrangian approach\footnote{Our choice of strong and electromagnetic interaction Lagrangians can be found in Ref.~\cite{RPRlambda}}. As a result, we need to introduce only three parameters
\begin{equation}\label{eq:parameters}
 g_{K^+\Sigma^0p}\, , \quad G^{v,t}_{K^{\ast+}\Sigma^0p} = \kappa_{K^{\ast+}K^+} \dfrac{e\,g^{v,t}_{K^{\ast+}\Sigma^0p}}{4\pi} \, ,
\end{equation}
with $g_{K^+\Sigma^0p}$, $g^v_{K^{\ast+}\Sigma^0p}$ and $g^t_{K^{\ast+}\Sigma^0p}$ the coupling constants at the strong interaction vertex and $\kappa_{K^{\ast+}K^+}$ the $K^{\ast+}(892)$'s transition magnetic moment. To obtain a Reggeized amplitude, which effectively incorporates the transfer of an entire trajectory, one substitutes the Feynman $(t-m^2_{K^{(\ast)+}})^{-1}$ propagator in the $t$-channel Born diagram by the corresponding Regge propagator
\begin{equation}\begin{split} \label{eq: reggeprop}
\mathcal{P}^{K^+(494)}_{\text{Regge}}(s,t) &= \left(\dfrac{s}{s_0}\right)^{\alpha_{K^+}(t)}
\dfrac{e^{-i\pi\alpha_{K^+}(t)}}{\sin\bigl(\pi\alpha_{K^+}(t)\bigr)} \; \dfrac{\pi \alpha'_{K^+}}{\Gamma\bigl(1+\alpha_{K^+}(t)\bigr)}  \,,
\\
\mathcal{P}^{K^{\ast +}(892)}_{\text{Regge}}(s,t) &= \left(\dfrac{s}{s_0}\right)^{\alpha_{K^{\ast+}}(t)-1} 
\dfrac{1}{\sin\bigl(\pi\alpha_{K^{\ast+}}(t)\bigr)} \; \dfrac{\pi \alpha'_{K^{\ast+}}}{\Gamma\bigl(\alpha_{K^{\ast+}}(t)\bigr)} \,,
\end{split}\end{equation}
with $s_0=1\,\text{GeV}^2$, $\alpha_{K^+}(t) = 0.70 \ (t-m_{K^+}^2)$ and $\alpha_{K^{\ast+}}(t) = 1 + 0.85 \ (t-m_{K^{\ast+}}^2)$, when $t$ and $m_{K^{(\ast)+}}^2$ are expressed in units of $\text{GeV}^2$. The data indicate that the trajectories are strongly degenerate. Consequently, the Regge propagators have either a constant or rotating phase. This phase cannot be deduced from first principles and needs to be determined by data.

In order to relate $n(\gamma,K^+)\Sigma^-$ to $p(\gamma,K^+)\Sigma^0$, it suffices to convert the coupling constants at the strong interaction vertex. Assuming isospin symmetry to be exact, the hadronic couplings are proportional to Clebsch-Gordan coefficients. One can find the following relations~\cite{RPRneutron}:
\begin{equation}\label{eq:convert}
 g_{K^{+}\Sigma^-n} = \sqrt{2}\,g_{K^{+}\Sigma^0p}\, , \quad 
 g^{v,t}_{K^{\ast+}\Sigma^-n} = \sqrt{2}\,g^{v,t}_{K^{\ast+}\Sigma^0p}\, .
\end{equation}
For the $p(\gamma,K^+)\Sigma^0$ reaction, the amplitude is made gauge-invariant by adding the electric part of the $s$-channel Born diagram. For $n(\gamma,K^+)\Sigma^-$, the same is achieved through the electric part of the $u$-channel Born diagram.

At sufficiently high energies ($\omega_{\text{lab}}\gtrsim4\,\text{GeV}$), a limited amount of $p(\gamma,K^+)\Sigma^0$ data is available, comprising differential cross sections~\cite{Boyarski} in addition to photon-beam asymmetries~\cite{Quinn}. These data show no resonant features and are used to constrain the three parameters of Eq.~\eqref{eq:parameters}. In Ref.~\cite{RPRsigma}, we identified a rotating and a constant phase for the $K^+(494)$ and $K^{\ast+}(892)$ trajectories respectively, as the only solution compatible with the data. The available 57 data points, did not allow us to single out a unique parametrization in that the sign of $G^{t}_{K^{\ast+}\Sigma^0p}$ remains undetermined~\cite{RPRsigma}. The two model variants, that yield the best description of the high-energy data, were labeled Regge-3 and Regge-4. The Regge model's amplitude can be interpreted as the asymptotic form of the full amplitude for large $s$ and small $|t|$. By extrapolating it into the so-called resonance region ($W\lesssim2.5\,\text{GeV}$), we can try to evaluate down to what energies the simple Regge model holds.

\begin{figure}
 \centering
 \includegraphics[width=.86\textwidth]{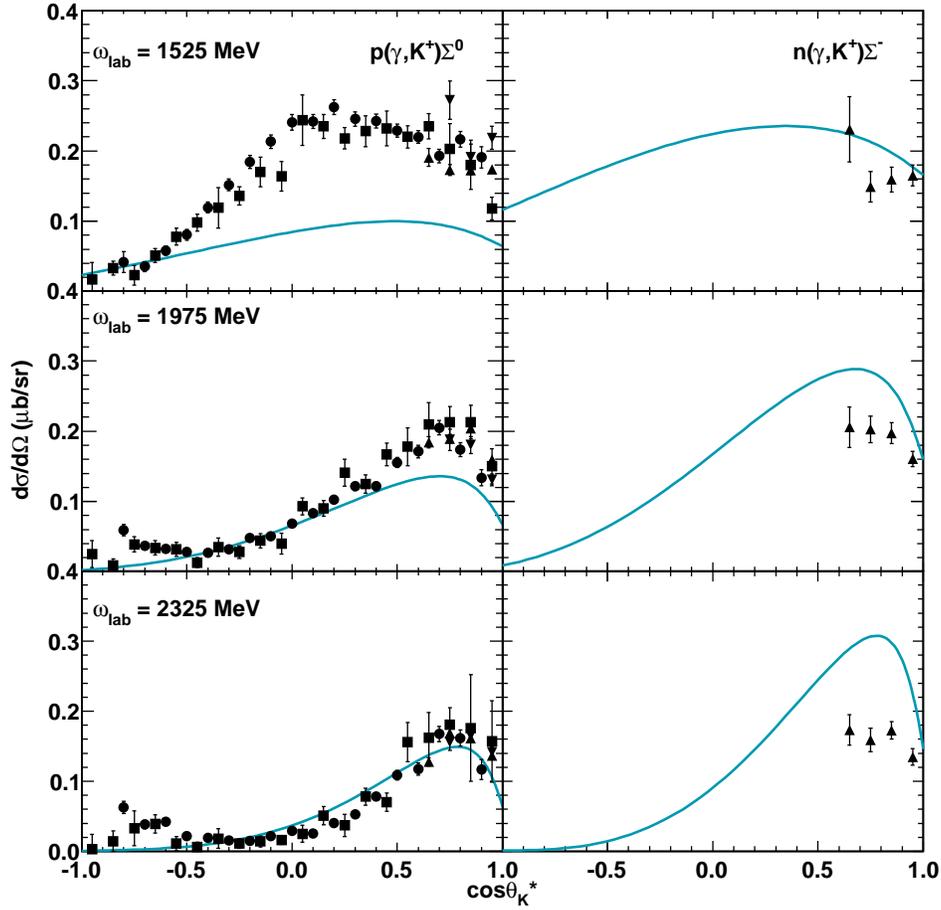}
  \caption{The differential cross section for $p(\gamma,K^+)\Sigma^0$~(left panels) and $n(\gamma,K^+)\Sigma^-$~(right panels) as a function of the kaon's center-of-mass scattering angle. The solid curve are calculations with the Regge-3 model. Data from Refs.~\cite{SAPHIR03}~($\blacksquare$), \cite{CLASdcs06}~($\bullet$), \cite{LEPSiso6}~($\blacktriangle$) and \cite{LEPSSumihama}~($\blacktriangledown$).}
 \label{fig:dcs}
\end{figure}

\begin{figure}
 \centering
 \includegraphics[width=.86\textwidth]{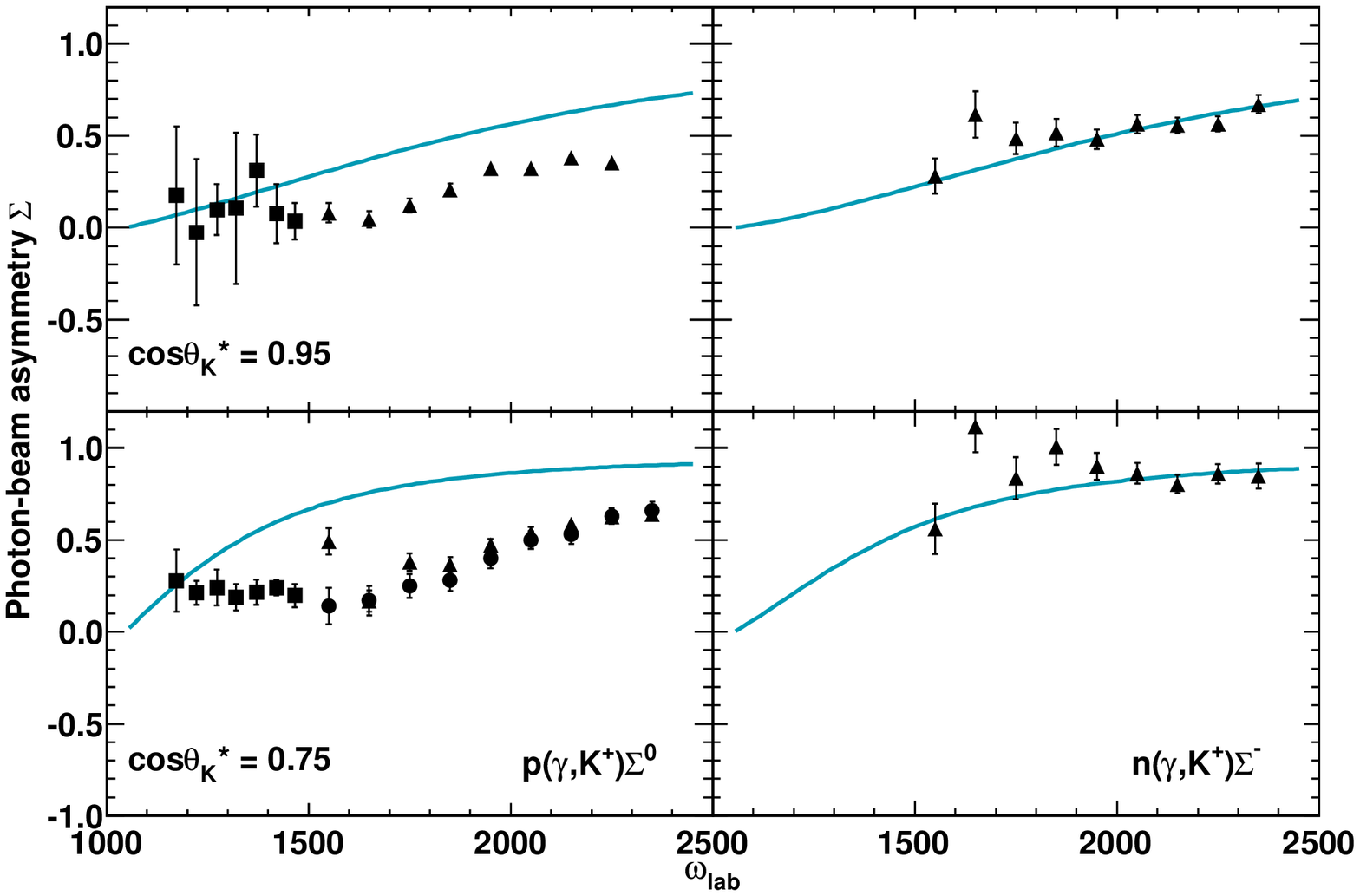}
  \caption{The photon-beam asymmetry for $p(\gamma,K^+)\Sigma^0$~(left panels) and $n(\gamma,K^+)\Sigma^-$~(right panels) as a function of the incoming photon's lab energy. The solid curve are calculations with the Regge-3 model. Data from Refs.~\cite{LEPSiso6}~($\blacktriangle$),\cite{GRAALrecpho}~($\blacksquare$) and \cite{LEPSpho}~($\bullet$).}
 \label{fig:pho}
\end{figure}

Figure~\ref{fig:dcs} features the angular dependence of the $p(\gamma,K^+)\Sigma^0$ differential cross section in three representative energy bins. As expected, the Regge model performs best at higher energies. At $\omega_{\text{lab}}=1525\,\text{MeV}$, the cross section is significantly underestimated, hinting at the presence of resonances. In Refs.~\cite{RPRsigma,RPRlambda,RPRelectro}, we have shown that by adding resonant $s$-channel contributions to the Regge amplitude, one can considerably improve on the overall description of the data. We coined the resulting hybrid approach Regge-plus-resonance~(RPR). In figure~\ref{fig:pho}, we show the photon-beam asymmetry for two forward-angle cosine bins over an extended energy range. The Regge model predicts a vanishing asymmetry at threshold which steadily rises towards $+1$ as energy increases. This is confirmed by the data, although the measured asymmetry is smaller. 

Figures~\ref{fig:dcs} and \ref{fig:pho}, also contain predictions of the $n(\gamma,K^+)\Sigma^-$ observables. On the basis of Eq.~\eqref{eq:convert}, one would expect the $n(\gamma,K^+)\Sigma^-$ cross section to be roughly twice as large as for $p(\gamma,K^+)\Sigma^0$. The scarce neutron data rather hints at cross sections of equal magnitude. The photon-beam asymmetry is nicely reproduced.

In summary, we have presented a simple Regge approach to electromagnetic kaon production from the proton and neutron. The neutron results are anchored to the proton ones through isospin symmetry. The mere three free parameters in our approach are fitted to the 57 available high-energy data points for $p(\gamma,K^+)\Sigma^0$. Owing to the $t$-channel dominance and the absence of a prevailing resonance, this model can account for the gross features of both the $p(\gamma,K^+)\Sigma^0$ and $n(\gamma,K^+)\Sigma^-$ differential cross sections and photon-beam asymmetries within the resonance region. This result corroborates earlier findings~\cite{RPRsigma,RPRelectro,GuidalUpdate}.

\section{Acknowledgments} 
This work was supported by the Research Foundation -- Flanders (FWO) and the research council of Ghent University.

\bibliographystyle{aipproc}
\bibliography{bibliography}

\begin{thebibliography}{15}
\expandafter\ifx\csname natexlab\endcsname\relax\def\natexlab#1{#1}\fi
\providecommand{\enquote}[1]{``#1''}
\expandafter\ifx\csname url\endcsname\relax
  \def\url#1{\texttt{#1}}\fi
\expandafter\ifx\csname urlprefix\endcsname\relax\def\urlprefix{URL }\fi
\providecommand{\eprint}[2][]{\url{#2}}

\bibitem[Guidal et~al.(1997)]{GuidalPhotoProdKandPi}
M.~Guidal, J.~M. Laget, and M.~Vanderhaeghen, \emph{Nucl. Phys.} \textbf{A627},
  645 (1997).

\bibitem[Guidal(1997)]{PHDguidal}
M.~Guidal, Ph.D. thesis, {Universit\'e de Paris-Sud, U.F.R. Scientifique
  d'Orsay} (1997).

\bibitem[Corthals et~al.(2007{\natexlab{a}})]{RPRsigma}
T.~Corthals, D.~G. Ireland, T.~Van~Cauteren, and J.~Ryckebusch, \emph{Phys.
  Rev.} \textbf{C75}, 045204 (2007{\natexlab{a}}).

\bibitem[Corthals et~al.(2006)]{RPRlambda}
T.~Corthals, J.~Ryckebusch, and T.~Van~Cauteren, \emph{Phys. Rev.}
  \textbf{C73}, 045207 (2006).

\bibitem[Vancraeyveld et~al.(2009)]{RPRneutron}
P.~Vancraeyveld, L.~De~Cruz, J.~Ryckebusch, and T.~Van~Cauteren  (2009), {in
  preparation}.

\bibitem[Boyarski et~al.({1969})]{Boyarski}
A.~M. Boyarski, et~al., \emph{{Phys.~Rev.~Lett.}} \textbf{{22}}, {1131}
  ({1969}).

\bibitem[Quinn et~al.(1979)]{Quinn}
D.~J. Quinn, et~al., \emph{{Phys.~Rev.}} \textbf{D20}, 1553 (1979).

\bibitem[Glander et~al.(2004)]{SAPHIR03}
K.~H. Glander, et~al., \emph{Eur. Phys. J.} \textbf{A19}, 251 (2004).

\bibitem[Bradford et~al.(2006)]{CLASdcs06}
R.~Bradford, et~al., \emph{Phys. Rev.} \textbf{C73}, 035202 (2006).

\bibitem[Kohri et~al.(2006)]{LEPSiso6}
H.~Kohri, et~al., \emph{Phys. Rev. Lett.} \textbf{97}, 082003 (2006).

\bibitem[Sumihama et~al.(2006)]{LEPSSumihama}
M.~Sumihama, et~al., \emph{Phys. Rev.} \textbf{C73}, 035214 (2006).

\bibitem[Ll{\`e}res et~al.(2007)]{GRAALrecpho}
A.~Ll{\`e}res, et~al., \emph{Eur. Phys. J.} \textbf{A31}, 79 (2007).

\bibitem[Zegers et~al.(2003)]{LEPSpho}
R.~G.~T. Zegers, et~al., \emph{Phys. Rev. Lett.} \textbf{91}, 092001 (2003).

\bibitem[Corthals et~al.(2007{\natexlab{b}})]{RPRelectro}
T.~Corthals, T.~Van~Cauteren, P.~Vancraeyveld, J.~Ryckebusch, and D.~G.
  Ireland, \emph{Phys. Lett.} \textbf{B656}, 186 (2007{\natexlab{b}}).

\bibitem[Guidal et~al.(2003)]{GuidalUpdate}
M.~Guidal, J.~M. Laget, and M.~Vanderhaeghen, \emph{Phys. Rev.} \textbf{C68},
  058201 (2003).

\end{thebibliography}

\end{document}